# BALLISTIC ELECTRON TRANSPORT IN WRINKLED SUPERLATTICES

T.L. Mitran, G.A. Nemnes, L. Ion and Daniela Dragoman

(tudor@solid.fizica.unibuc.ro, nemnes@solid.fizica.unibuc.ro, lucian@solid.fizica.unibuc.ro, danieladragoman@yahoo.com)

University of Bucharest, Faculty of Physics,

Materials and Devices for Electronics and Optoelectronics Research Center,

P.O. Box MG-11, 077125 Magurele-Ilfov, Romania

**Abstract**

Inspired by the problem of elastic wave scattering on wrinkled interfaces, we studied the scattering of ballistic electrons on a wrinkled potential energy region. The electron transmission coefficient depends on both wrinkle amplitude and periodicity, having different behaviors for positive and negative scattering potential energies. For scattering on potential barriers, minibands appear in electron transmission, as in superlattices, whereas for scattering on periodic potential wells the transmission coefficient has a more complex form. Besides suggesting that tuning of electron transmission is possible by modifying the scattering potential via voltages on wrinkled gate electrodes, our results emphasize the analogies between ballistic electrons and elastic waves even in scattering problems on non-typical configurations.

# I. INTRODUCTION

Devices that are able to precisely control charge transport have been one of the main goals of solid state physics, and, presently, of nanoelectronics. One of the methods to accomplish this is the fabrication of nanosize artificial structures, such as superlattices. The engineering of these artificial structures can be paralleled to the development of photonic crystals in optics and phononic crystals in periodic solid state materials, which evolved due to the need to control the propagation of electromagnetic and, respectively, elastic or heat waves at given frequencies. Indeed, due to the well-known analogies between ballistic electrons and light waves [1, 2], between electrons and phonons [3] or acoustic waves [4], and between phonons and electromagnetic waves (see, for example, [5] and the references therein), almost any conceptual development or application in one of these domains, i.e. nanoelectronics, optics or phononics, can be adapted to the others. Among many examples, we mention here (besides the emergence of photonic and phononic crystals in analogy to crystalline solids) only the universal conduction fluctuations of light [6], the photonic classical [7] and quantum [8] Hall effects, the photonic [9] and acoustic spin Hall effect [10], and the observation of graphene-like Dirac points in photonic crystals [11].

In this paper, we study an electronic analog of a recently proposed method to control the elastic wave propagation in layered media with interfacial wrinkling [12]. In particular, we are interested in the effect of tunable wrinkled scattering potentials on the transport properties, especially on the transmission coefficient, of two-dimensional semiconductors. The investigated systems consist of a rectangular scattering region connected to two ideal leads, and subject to an oscillating scattering potential parallel to the direction of the leads. In principle, in this configuration, the potential energy, as well as the number, width and periodicity of wrinkles can be modified. Such a wrinkled scattering region can be implemented by a meandering gate, which transforms into a straight gate in the leads, in order to preserve the continuity of the system. The transmission function is studied in all cases for different values of scattering potentials and electron energies by using the scattering formalism of the R-matrix method.

It should be noted that although meandering gates have been used before in ion-sensitive field-effect transistors [13], in plasmonic THz detectors [14], to enhance the photoresponse at the second harmonic of the cyclotron resonance in a two-dimensional electron gas [15], or to maximize the active area of graphene-on-$MoS_2$ capacitors [16], no study on ballistic electron transport in such structures has been performed up to now. Our results suggest that by tuning the amplitude and period of the wrinkles, as well as by modifying the scattering potential, it is possible to adjust the transmission coefficient at the desired value. Such an adjustment is not possible for a straight potential barrier of the same width.

## II. THEORETICAL BACKGROUND

The R-matrix formalism was developed by Wigner and Eisenbud in 1947 as a nuclear scattering model [17], but has gained attraction as an efficient numerical means of simulating electron scattering in semiconductors physics [18, 19], being applied to the study of nanoscale transistors [20-22], the thermoelectrical properties of nanowires [23] and to spin transport [24]. For the present study, it was chosen as the simulation environment for quantum transport because of its numerical efficiency, which is also described in [25]. This numerical speed gain is obtained by separating the problem into an energy-independent part that solves the Wigner-Eisenbud eigenvalue problem and a second one, in which the transmission is computed at each energy value. Such a method is well suited for systems in which the transmission function varies rapidly with the energy, and has narrow peaks. The system is split into a scattering region, with an spatially varying potential in both $x$ and $y$ directions, and leads that act as source and drain, where the potential is varied only perpendicular to the direction of transport (Fig. 1). Electrons originating from the source (left lead) interact with the potential from the scattering region by obeying the stationary Schrödinger equation in the effective mass approximation,

$$H\Psi(\vec{r};E) = E\Psi(\vec{r};E),\qquad(1)$$

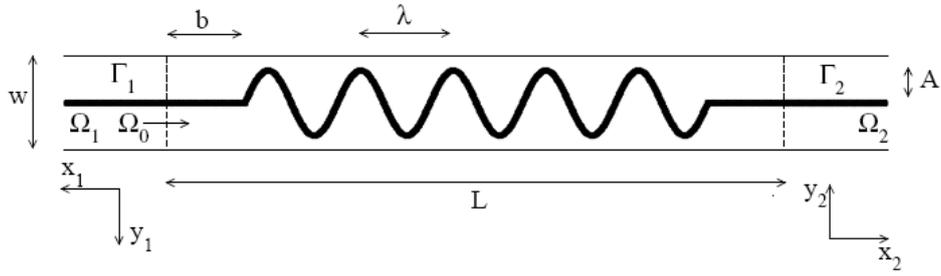

Fig. 1. Schematic representation of the scattering region and leads for the 2D system

before continuing through to the drain (right lead). The one-particle Hamiltonian in the effective mass approximation is

$$H = -\frac{\hbar^2}{2m^*}\Delta + W(\vec{r})\qquad(2)$$

where $m^*$ denotes the effective mass and $\vec{r} = (x, y)$ is the position vector.

As seen in Fig. 1, the scattering region contains an oscillating, or wrinkled, potential with a constant width of 1 nm that is continued in the translational invariant leads by a non-oscillating potential of identical width. The variable parameters of the wrinkled potential inside the scattering region are: the geometrical amplitude of the oscillation $A$, its period $\lambda$, and the (positive or negative) value of the applied potential, denoted by $V_0$. Throughout this paper we consider configurations containing 5 wrinkles, such that the length of the wrinkled potential region is $5\lambda$.

Inside the invariant leads, the solution of the Schrödinger equation has the form:

$$\Psi_s(\vec{r} \in \Omega_s; E) = \sum_i \Psi_\nu^{in} \exp(-ik_\nu x_s)\Phi_\nu(y_s) + \sum_i \Psi_\nu^{out} \exp(ik_\nu x_s)\Phi_\nu(y_s), \qquad (3)$$

where $\Psi_\nu^{in}$ and $\Psi_\nu^{out}$ are complex coefficients corresponding to the lead with index $s$. The composite index $\nu = (s, i)$ denotes the lead index $s$ and the channel index $i$. The coordinate system was chosen such that the $x$ axis points away from the scattering region. The scattering boundary conditions for an electron in channel $\nu$ are found by setting $\Psi_{\nu'}^{in} = 1/\sqrt{2\pi}\delta_{\nu\nu'}$. The wavevectors along the leads are expressed as $k_\nu = \hbar^{-1}\sqrt{2m^*(E - E_\perp^\nu)}$ and $E_\perp^\nu$ are the energies of the perpendicular modes obtained from the one-dimensional Schrödinger equation. The complex coefficients $\Psi_\nu^{in}$, $\Psi_\nu^{out}$ are related through the scattering S-matrix by $\vec{\Psi}^{out} = S\vec{\Psi}^{in}$. The S-matrix can also be used to obtain the total transmission between the source and drain: $T_{ss'}(E) = \sum_{i,i'} |\tilde{S}_{\nu\nu'}(E)|^2$, where $\tilde{S} = k^{1/2}Sk^{-1/2}$ and the summation is only over open channels. In the R-matrix formalism, the S-matrix can be written as:

$$S = -\frac{1 + \frac{i}{m^*}Rk}{1 - \frac{i}{m^*}Rk} \qquad (4)$$

where $(k)_{\nu,\nu'} = k_\nu \delta_{\nu\nu'}$ and the R-matrix is expressed as

$$(R)_{vv'}(E) = -\frac{\hbar^2}{2} \sum_{l=0}^{\infty} \frac{(\chi_l)_v (\chi_l^*)_{v'}}{E - \epsilon_l}. \tag{5}$$

with

$$(\chi_l)_v = \int_{\Gamma_s} d\Gamma_s \Phi_v(y_s) \chi_l(\vec{r} \in \Gamma_s), \tag{6}$$

where $\chi_l(\vec{r} \in \Omega_0)$ and $\epsilon_l$ are the Wigner-Eisenbud functions and energies, and $\Omega_0$ represents the scattering region. These are the eigenfunctions and eigenvalues of the Hamiltonian in eq. (1) with new and fixed boundary conditions on the boundary of the scattering region with lead $s$, or $\Gamma_s$:

$$\left[\frac{\partial \chi_l}{\partial x_s}\right]_{\Gamma_s} = 0. \tag{7}$$

It is important to note that even though the determination of the Wigner-Eisenbud energies and functions ($\chi_l$, $\epsilon_l$) and the overlap integrals $(\chi_l)_v$ is the most time consuming step, it is only done one time since it is energy independent. After that, the R-matrix can be computed for each energy from eq. (5) by performing a summation over the Wigner-Eisenbud indexes. The S-matrix is determined by using eq. (4) and finding the inverse of the $[1 - i/m^*Rk]$ matrix.

## III. RESULTS AND DISCUSSION

The system under investigation is a two-dimensional electron gas in the form of a nanoribbon, with a width $w = 20$ nm and length $L = 120$ nm, in which the effective electron mass is $m^* =$

0.0655 $m_0$. This effective mass value is characteristic for GaAs, which can form a two-dimensional electron gas at the interface with AlGaAs (see, for example, [26]). Note that $L$ is not the length of the wrinkled region (which equals $5\lambda$; see Fig. 1), but includes also the straight continuation of this region. The constant amplitude $V_0$ of the sinusoidal electrostatic potential can be varied by modifying the gate voltage, and is considered in our simulations to vary between –1 eV and 1 eV.

Using the R-matrix formalism, we have calculated the transmission coefficient as a function of the electron energy $E$ and potential $V_0$, for the cases when the amplitude $A$ and the period $\lambda$ of the wrinkles vary. The results are plotted in Fig. 2 for the case $A = 0$ (straight gate) and in Fig. 3 for $A$ = 3, 6, and 9 (upper, central and lower rows, respectively) and $\lambda$ = 16 nm, 18 nm and 20 nm (left, centre and right columns, respectively).

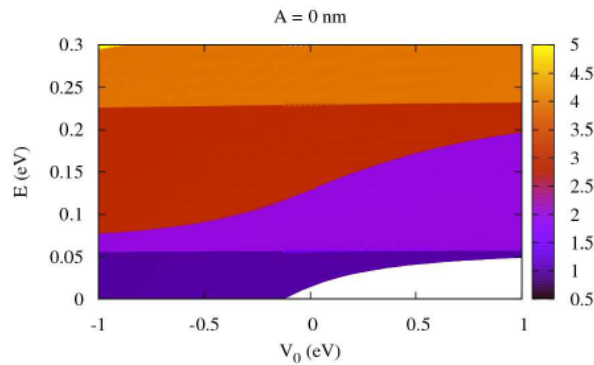

Fig. 2. Transmission coefficient in the case of straight scattering potentials ($A = 0$) as a function of electron energy ($E$) and scattering potential ($V_0$)

In all cases, in the region in the lower right corner of the figures, with white color, electron propagation is prohibited by the transversal constraint/quantum confinement, which imposes a minimum energy value for propagation in the first allowed mode. Note that this minimum energy depends on the confining potential; it increases as $V_0$ increases and vanishes at $V_0$ values smaller than about –0.12 eV when the transverse spatial confinement of the electron propagation in the nanoribbon is no longer apparent. Indeed, the application of a potential energy on the 1-nm-wide straight continuation of the wrinkled region inside the leads modifies the effective nanoribbon width; the nanoribbon is effectively wider for negative $V_0$ values and effectively narrower for positive values of this parameter.

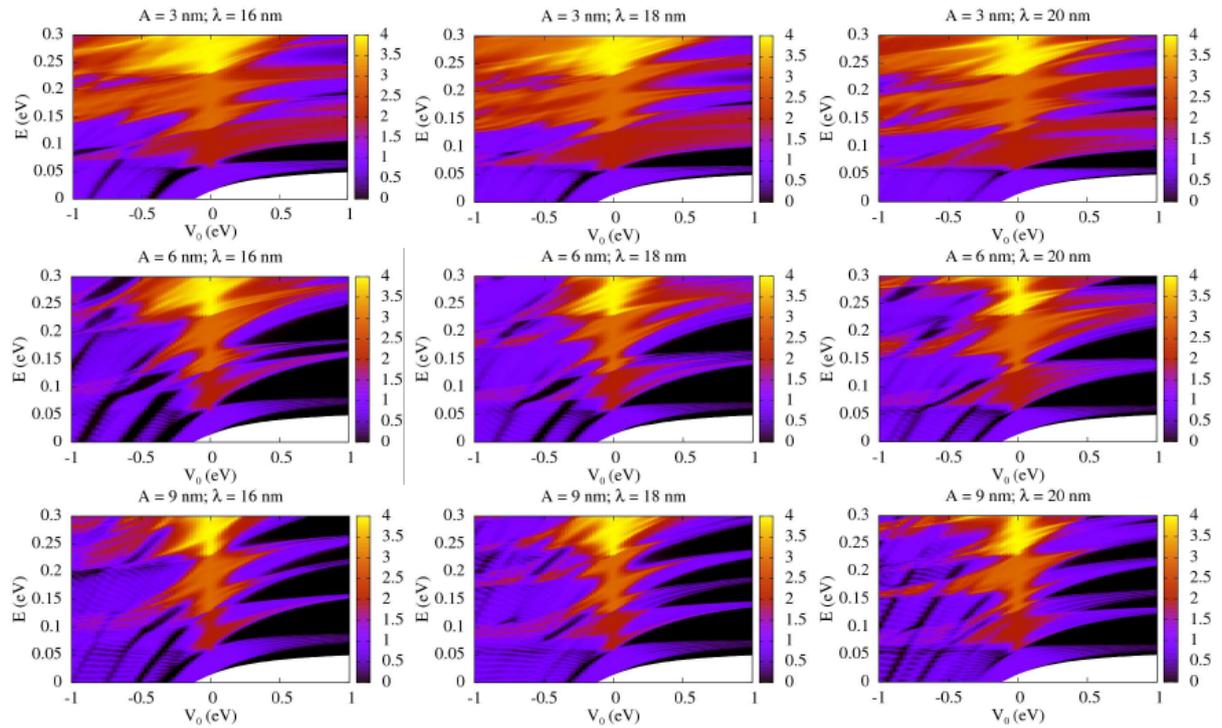

Fig. 3. Transmission coefficient as a function of electron energy ($E$) and scattering potential ($V_0$) for different combinations of amplitudes ($A$) and periods ($\lambda$) of the scattering potential

From Figs. 2 and 3 it can be seen that the transmission coefficient has a totally different behavior for the cases of a straight and a wrinkled scattering potential. In the first case the transmission increases uniformly with the number of open channels, as expected. On the other hand, for a wrinkled scattering potential the transmission coefficient has a complex dependence on $E$ and $V_0$, with a clear difference in behavior for negative and positive values of $V_0$. The behavior of the transmission coefficient for positive $V_0$ can be easily understood in terms of appearance of minibands in the structure containing periodic potential barriers. The width of the allowed minibands decreases and their central position is shifted towards higher energies as the barrier height increases [27]. The same appearance of bandgaps upon wrinkling initiations was observed also in the case of elastic waves [12]. A positive $V_0$ for electrons seems thus to be correlated with a higher stiffness of the deformable interfacial layers in [12].

The application of a scattering potential, which induces a transverse non-homogeneous potential energy distribution, does not modify just the transmission coefficient, but also the wavefunction in the leads. Examples of wavefunction propagation, more precisely of the non-normalized probability distribution, in the same wrinkle configuration, for the fundamental mode at two different electron energies, are represented in Figs. 4 and 5 for the situation when $V_0 = -0.5$ eV and 0.5 eV, respectively. These figures illustrate the effect of the narrow and straight continuation of the wrinkled region on the form of the fundamental mode in the leads. For the same energy values of incident electrons, the form of the wavefunction depends on whether the applied potential is attractive (negative $V_0$ values) or repulsive (positive $V_0$); the applied potential changes the form of wavefunctions since it is not applied on the whole width of the nanoribbon but only in its central part.

Also, the wrinkled scattering potential influences the form (not only the *y*-dependence of the amplitude, but also of the phase) of the transmitted wavefunction. For example, in the upper Fig. 4 constructive interference/wavefunction maxima seems to occur near the center of the nanoribbon before the scattering region, while after scattering destructive interference is observed close to the nanoribbon center. Again, in the upper Fig. 5, the transmitted wavefunction, while having the same general form, has a different oscillation period along the *x* axis.

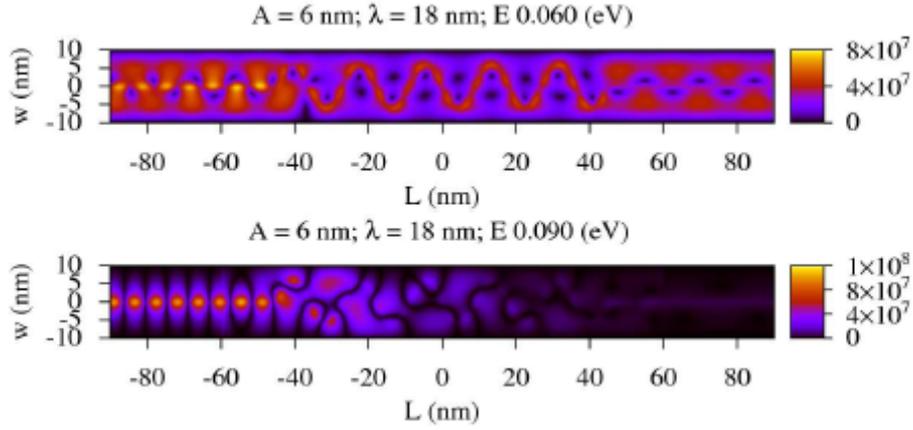

Fig. 4. Wavefunction representation for an attractive scattering potential

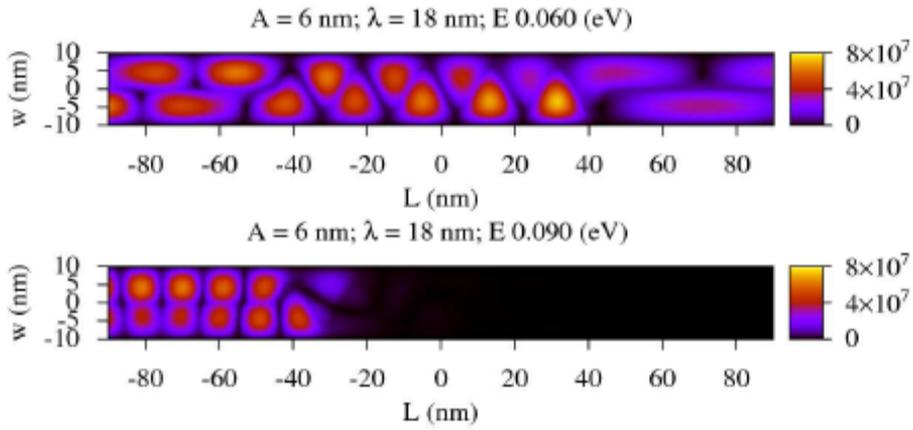

Fig. 5. Wavefunction representation for a repulsive scattering potential

The effect on the propagation of elastic waves of the periodic structure with smaller stiffness of the deformable interfacial layers was not studied in [12]. This situation could correspond to a negative $V_0$ value, for which the dependence of electron transmission on $E$ and $V_0$ is quite complex, as seen from Fig. 3. The periodic scattering structure with wells instead of barriers is characterized by interferences not only between quantum wavefunctions scattered of adjacent wells but also by quantum interferences in the wells. The superposition of all such interferences, with different origins, forms the complex pattern in Fig. 3.

To better understand the differences between the effect of periodic barriers and wells on quantum wavefunction scattering in a nanoribbon we performed a simpler simulation of a corresponding one-dimensional scattering problem, in which the wrinkled scattering region was replaced by stripes with the same width and position as encountered by electrons propagating at different *y* values in Fig. 1. The one-dimensional configuration is represented in Fig. 6(a), where the gray line illustrates the wrinkled region and its straight continuation in the left lead, the vertical thick solid (dashed) black lines represent the positions of scattering regions for electrons propagating along trajectories for which *y* = 0 ( *y* ≠ 0 ); these trajectories are illustrated with thin horizontal black lines of the same type. Note that the width of the scattering regions for electrons with *y* ≠ 0 is wider than for electrons propagating along *y* = 0, and the position of the scattering regions is not equidistant in a period *λ*, as is the case for *y* = 0.

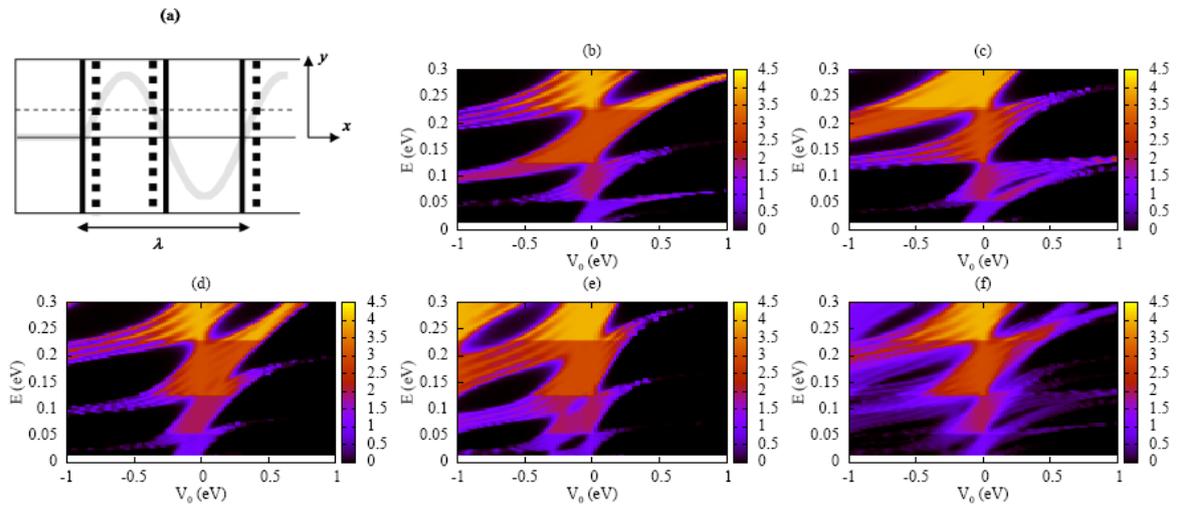

Fig. 6. Schematic representation of the 1D system (a) and the transmission as a function of electron energy (*E*) and scattering potential (*V*$_0$) for trajectories along *y* = 0 (b), 0.33*A* (c), 0.66*A* (d) and 0.85*A* (e), and the mean value of (b)-(e) in (f)

The resulting transmission coefficients for electrons with trajectories along $y = 0$, $0.33A$, $0.66A$ and $0.85A$ are displayed in Figs. 6(b), 6(c), 6(d) and 6(e), respectively. In all cases, since in the one-dimensional problem the continuation of the wrinkled region in the leads was not taken into account, the white color region representing prohibited electron propagation by the transversal constraint/quantum confinement has the same width, irrespective of the $V_0$ potential value applied on the transverse scattering region. The width of the white-color-region is the same as that in Figs. 3 for $V_0 = 0$, i.e. for a lead with a homogeneous (and zero valued) potential distribution.

From Fig. 6(b) it follows also that the transmission coefficient $T$ has similar minibands for positive and negative $V_0$ values. However, the effect of barriers and wells (positive and negative values of the scattering potential) on electron propagation is more dissimilar for increasing $y$ values, for which the effective width of the scattering potential region. In particular, for wide enough wells distinct interferences form inside them, which superimpose on the interferences due to adjacent wells, the result being the complex interference pattern in Fig. 6(e). In this figure, the transmission for positive $V_0$ is inhibited by the wide barriers, while for negative $V_0$ values the wide transmission bands do not monotonously decrease in width with the applied potential (as in Figs. 6(b)-6(d)) because of the influence of constructive interferences forming inside the wide-enough wells.

This simple one-dimensional model can provide a helpful insight in the behavior of the transmission coefficient of the two-dimensional problem. More precisely, assuming that the total transmission coefficient is just a sum of the contributions for different $y$ values, the form of the mean value of $T$ in Figs. 6(b)-6(d), represented in Fig. 6(f), is similar to the dependence of $T$ on $E$ and $V_0$ in the two-dimensional problem. This similarity is quite remarkable taking into account the drastic simplification of the scattering geometry. Of course, the simple one-dimensional cannot capture all the details of the two-dimensional scattering problem, in particular the dependence of the white region on the value of the scattering potential.

## IV. CONCLUSIONS

In summary, inspired by a similar problem for elastic wave scattering, we have studied the scattering of ballistic electrons on a wrinkled potential energy region. The transmission coefficient behavior on the electron energy and the amplitude of the potential depends on both wrinkle amplitude and periodicity. For scattering on periodic potential barriers, minibands appear in electron transmission, as in any superlattice, whereas for scattering on periodic potential wells the transmission coefficient has a more complex form. For a scattering potential/gate electrode of a specific form, the value of the transmission coefficient can be tuned to a desired value by modifying the applied gate voltage. Such tuning is not possible in a structure with a straight scattering region with the same width.

Apart from studying the effect on electron transmission of a wrinkled scattering potential, our results emphasize once more the analogies between ballistic electrons and elastic waves. In this respect, the similarities between bandgap formation in wrinkled structures have been observed for scattering on potential barriers for electrons and higher stiffness of interfacial layers for elastic waves. The situation in optics would correspond to scattering on regions with a smaller refractive index. The opposite situation, not considered for the elastic wave case, has revealed a complex behavior for ballistic electrons. The corresponding situation in optics, that of scattering on regions with a higher refractive index, although not studied in a configuration similar to Fig. 1, at least to the authors' knowledge, would imply anti-resonant reflections, as in ARROW-like structures (see [28], for example).